\documentclass[prd,notitlepage,longbibliography,nofootinbib,superscriptaddress,onecolumn,preprintnumbers]{revtex4-2}
\usepackage[utf8]{inputenc}

\usepackage{bm}
\usepackage{comment} 
\usepackage[colorlinks=true,urlcolor=blue,anchorcolor=black,citecolor=blue,linkcolor=red,filecolor=black,menucolor=black,pagecolor=black,linktocpage=true,pdfproducer=medialab,pdfa=true]{hyperref}
\usepackage{graphicx}
\usepackage{amsmath,latexsym,amssymb,mathrsfs,ascmac,mathtools}
\usepackage{multirow}
\usepackage{braket}
\usepackage{placeins}
\begin{document}

\title{Photon Surfaces in Higher-Curvature Gravity: Implications for Quasinormal Modes and Gravitational Lensing}

\author{Takamasa Kanai}
\email{kanai@kochi-ct.ac.jp}

\affiliation{Department of Social Design Engineering,
National Institute of Technology (KOSEN), Kochi College,
200-1 Monobe Otsu, Nankoku, Kochi, 783-8508, Japan}

\begin{abstract}
Effective field theory (EFT) provides a systematic framework to describe possible deviations from general relativity through higher-curvature corrections to the gravitational action, capturing the low-energy effects of an underlying fundamental theory.

In this work, we investigate quasinormal modes (QNMs) of test fields, as well as weak and strong gravitational lensing, in static, spherically symmetric spacetimes with higher-curvature corrections, focusing on the behavior of null geodesics near the photon sphere. The QNM analysis is restricted to the test-field limit, in which fields propagate on the fixed background without backreaction, and gravitational perturbations of the metric are not considered. Adopting the strong deflection limit formalism developed by Bozza, we derive the logarithmic divergence structure of the deflection angle and explicitly separate the divergent and regular contributions.

The dimensionful couplings are normalized by appropriate powers of the horizon radius $r_s = 2M$ to define dimensionless expansion parameters, and our perturbative treatment is valid when these parameters are small, i.e., when the curvature scale of the black hole remains below the EFT cutoff. Within a simplified setup with $2M=1$, we analyze how deviations from general relativity, parametrized within this EFT framework, modify the photon sphere radius, the critical impact parameter, and the coefficients governing the strong deflection expansion. We show how these quantities reflect the effects of higher-curvature corrections to the gravitational action.

Our results provide a systematic characterization of the effects of higher-curvature EFT corrections on null geodesics, gravitational lensing, and test-field QNMs. In particular, we demonstrate how the corrections to the photon sphere and strong-deflection coefficients are related to the EFT couplings, and how the corresponding modifications appear in the eikonal behavior of test-field QNMs. These results clarify the connections between higher-curvature corrections to the gravitational action and characteristic properties of the surrounding spacetime geometry.
\end{abstract}
\maketitle

\section{Introduction}

Gravitational phenomena in the strong-field regime provide a unique opportunity to test the nature of gravity beyond the weak-field and post-Newtonian approximations. In particular, null geodesics propagating near compact objects such as black holes exhibit characteristic features associated with unstable photon orbits. Photon surfaces provide a general geometric framework for characterizing certain families of null geodesics \cite{Claudel:2000yi}. In static, spherically symmetric spacetimes, the relevant structure for strong gravitational lensing and black hole shadows is the unstable photon sphere \cite{Virbhadra:1999nm,Claudel:2000yi}. Its properties are closely related to observable features of black hole shadows, which have become accessible through horizon-scale observations \cite{Falcke:1999pj,EventHorizonTelescope:2019dse,EventHorizonTelescope:2022wkp,Vagnozzi:2022moj}, and to the deflection of light in the strong-field regime \cite{Virbhadra:1999nm,Bozza:2001xd,Bozza:2002zj,Gibbons:2008rj}. Unstable photon orbits are also relevant to the eikonal limit of quasinormal modes (QNMs) of suitable test fields, although the correspondence depends on the type of perturbation and the gravitational theory under consideration \cite{Chandrasekhar:1975zza,Iyer:1986np,Kokkotas:1999bd,Buonanno:2006ui}.

Among these phenomena, strong gravitational lensing provides a valuable probe of the geometry around compact objects. When light rays approach the unstable photon sphere, the deflection angle grows without bound as the impact parameter approaches its critical value. In the strong deflection limit formalism developed by Bozza \cite{Bozza:2002zj}, the deflection angle is expressed as the sum of a logarithmically divergent term and a regular contribution. The corresponding coefficients are determined by the metric functions and their derivatives evaluated at the photon sphere, making this formalism useful for investigating how modifications of the spacetime geometry affect strong gravitational lensing.

Effective field theory (EFT) provides a systematic framework for describing gravitational interactions and parametrizing deviations from general relativity through higher-curvature operators \cite{Weinberg:1978kz,Donoghue:1993eb,Donoghue:1994dn,Burgess:2003jk}. Such corrections can modify the background geometry and, depending on the couplings involved, the propagation of different types of perturbations. Consequently, geometric quantities associated with unstable null geodesics, including the photon sphere radius and the critical impact parameter, can acquire corrections determined by the EFT couplings. These modifications can in turn affect strong-field gravitational lensing and black hole shadow properties.

In particular, the logarithmic divergence and the regular contribution to the deflection angle in the strong deflection limit are characterized by coefficients determined by the local geometry near the photon sphere. Their corrections therefore provide a systematic characterization of the effects of higher-curvature operators on strong gravitational lensing. Whether such corrections can be constrained observationally depends on their magnitude and on the precision of current and future measurements.

The eikonal limit of QNMs provides another means of studying the dynamics of perturbations in black hole spacetimes. In general relativity, the eikonal QNMs of suitable test fields are closely related to the properties of unstable circular null geodesics. However, in higher-curvature theories, the propagation of gravitational perturbations need not be governed by the same effective geometry as that of photons, and the correspondence between QNMs and photon-sphere properties cannot in general be extended directly to gravitational perturbations. QNMs in higher-curvature and EFT-modified gravity theories have been investigated in Refs.~\cite{Antoniou:2024jku,Antoniou:2026nhh} and the references therein. In the present work, we use the eikonal QNMs of a test scalar field as a complementary characterization of the background geometry, without assuming that they directly describe the QNMs of gravitational perturbations.

Higher-curvature corrections to the Schwarzschild geometry and their effects on black-hole perturbations have been studied extensively in effective field theories of gravity. In particular, the leading corrections to the Schwarzschild solution and the corresponding gravitational QNMs were investigated in Ref.~\cite{Cardoso:2018ptl}, while the effects of cubic curvature corrections on black-hole perturbations and QNMs were further studied in Ref.~\cite{deRham:2020ejn}. These works provide the relevant background for the corrected Schwarzschild geometry and the QNM analysis considered here. More recently, higher-derivative curvature corrections to gravitational lensing have also been investigated. In our previous work~\cite{Kanai:2026qhi}, we studied strong gravitational lensing in EFT-corrected Reissner-Nordström spacetimes and demonstrated that higher-derivative corrections can modify the lensing properties in a characteristic manner.

Building on these studies, the present work investigates cubic and quartic curvature corrections in static, spherically symmetric spacetimes, with particular emphasis on the strong deflection regime. The corrected Schwarzschild geometry itself is not new; rather, we focus on the strong-deflection properties that follow from the higher-curvature corrections. In particular, we derive the corrections to the photon sphere radius and the critical impact parameter, and systematically calculate the corresponding strong-deflection coefficients. We also examine the eikonal QNMs of a test scalar field and compare their corrections with those of the lensing quantities. This allows us to characterize how different quantities associated with the background geometry respond to higher-curvature EFT corrections.

Our goal is to clarify how the coefficients in the strong deflection expansion encode the local geometry near the photon sphere and how they are modified by higher-curvature corrections within an EFT description. In particular, we investigate the dependence of these quantities on the EFT couplings and assess the extent to which their corrections may provide useful signatures of higher-curvature gravitational effects.

This paper is organized as follows. In Sec.~\ref{sec:photon_surface}, we review the concept of photon surfaces and their key properties in static spacetimes. In Sec.~\ref{sec.EFT corrections}, we introduce the effective field theory (EFT) framework for gravity adopted in this work. In Sec.~\ref{sec.photon surface EFT}, we construct the perturbative solution around the Schwarzschild background including higher-curvature corrections. Moreover, we determine the location of the photon surface and the critical impact parameter in the corrected spacetime. In Sec.~\ref{sec:qnm}, we compute the quasinormal mode (QNM) frequencies in the eikonal (large angular momentum) limit as an observable quantity. In Sec.~\ref{sec:weak lens}, we analyze gravitational lensing in the weak-field regime, whereas in Sec.~\ref{sec:strong lens}, we consider the strong-field regime. Finally, we summarize our results and discuss future prospects in Sec.~\ref{sec:conclusion}.

In this paper, we set the Newton constant and the speed of light equal to unity.

\section{Photon Surface}
\label{sec:photon_surface}

In this section, we review the notion of a photon surface and its geometric properties, following the formulation of Claudel, Virbhadra and Ellis~\cite{Claudel:2000yi}. Photon surfaces characterize the behavior of null geodesics in strong gravitational fields and provide a useful geometric framework for studying the propagation of light. In static, spherically symmetric spacetimes, a particularly important case is the unstable photon sphere, which governs the properties of photon rings and plays a central role in strong gravitational lensing. The unstable photon sphere is also closely related to the eikonal quasinormal modes of test fields, providing a connection between the spacetime geometry and these observables. In this work, we therefore focus specifically on the static, unstable photon sphere rather than photon surfaces in general.

\subsection{Definition and geometric characterization}

A \textit{photon surface} $\mathcal{S}$ is defined as a timelike three-dimensional hypersurface in a spacetime $(M, g_{ab})$ such that any null geodesic initially tangent to $\mathcal{S}$ continues to remain within $\mathcal{S}$  \cite{Claudel:2000yi}. This concept generalizes the more familiar notion of a photon sphere: while a photon sphere requires spherical symmetry and a constant time-lapse function, a photon surface
is defined purely in terms of the confinement of null geodesics and may in principle be dynamical or non-spherically symmetric.

The canonical example is the surface $r = 3M$ in the Schwarzschild spacetime, where all photon rings are located and form a photon sphere by virtue of spherical symmetry. By contrast, in a Kerr spacetime, no photon surface exists because the radial location of null geodesics confined to $r = \mathrm{const.}$ surfaces depends on their angular momentum; the photon region is instead an extended region bounded by photon orbits on the inner and outer boundaries~\cite{Teo:2003ltt,Perlick:2004tq,Grenzebach:2014fha,Galtsov:2019bty}.

To characterize a photon surface geometrically, let $n^a$ be the unit spacelike normal
to $\mathcal{S}$. The induced metric and the extrinsic curvature of $\mathcal{S}$ are
defined by
\begin{align}
    h_{ab} &= g_{ab} - n_a n_b, \label{eq:induced_metric} \\
    K_{ab} &= h_a{}^c \nabla_c n_b = \frac{1}{2} \mathcal{L}_n h_{ab},
    \label{eq:extrinsic_curvature}
\end{align}
where $\mathsterling_n$ denotes the Lie derivative with respect to $n^a$. Three equivalent conditions for $\mathcal{S}$ to be a photon surface are known
\cite{Claudel:2000yi}:
\begin{enumerate}
    \item Every affine-parametrized null geodesic of the submanifold $\mathcal{S}$ is
    simultaneously a null geodesic of the ambient spacetime $M$.

    \item For any null vector $k^a$ tangent to $\mathcal{S}$,
    \begin{equation}
        K_{ab} k^a k^b = 0.
        \label{eq:photon_surface_null}
    \end{equation}

    \item The hypersurface $\mathcal{S}$ is \textit{umbilical}, i.e.,
    \begin{equation}
        K_{ab} \propto h_{ab}.
        \label{eq:umbilicaleq:umbilical}
    \end{equation}
\end{enumerate}
These three conditions are mutually equivalent. The third condition~\eqref{eq:umbilicaleq:umbilical}, the umbilical condition, is the most convenient for practical computations and will be referred to as the \textit{photon surface condition} hereafter.

The physical interpretation is transparent from the variational principle. Consider a null geodesic of $M$ whose path lies within $\mathcal{S}$.
The action $\mathcal{S} = \int \mathcal{L}\, d\lambda$ with $\mathcal{L} = \frac{1}{2} g_{ab} k^a k^b$ is stationary under arbitrary path
deformations, including those restricted to $\mathcal{S}$. Hence, the null geodesic is also a geodesic of the submanifold. Conversely, for $\mathcal{S}$ to be a photon surface, the action must be stationary under deformations in the normal direction $n^a$; the change in $\mathcal{L}$ under such a deformation is proportional to $K_{ab} k^a k^b$, which yields condition~\eqref{eq:photon_surface_null}.
The umbilical form~\eqref{eq:umbilicaleq:umbilical} is the most general tensor structure consistent with condition~\eqref{eq:photon_surface_null} for arbitrary $k^a$.

\subsection{Uniqueness theorems}

A fundamental question is whether the Schwarzschild spacetime is the unique asymptotically flat vacuum spacetime that admits a static photon surface. This question has been addressed through a sequence of uniqueness theorems.

The first uniqueness result was established by Cederbaum~\cite{Cederbaum:2014gva,Cederbaum:2015aha} and subsequently strengthened by Cederbaum and Galloway~\cite{Cederbaum:2015fra}. In these works, the photon surface is taken to be a \textit{photon sphere}, defined as a static photon surface on which the time-lapse function $N = \sqrt{-g_{tt}}$ is constant. Analyzing the Einstein equations in the exterior region with appropriate boundary conditions on the photon sphere, one can show that the spacetime must be spherically symmetric, and therefore isometric to the Schwarzschild solution. This result has been generalized to electrovacuum spacetimes~\cite{Yazadjiev:2015jza,Cederbaum:2015fra}, to spacetimes with a conformal scalar field~\cite{Tomikawa:2016dqz,Tomikawa:2017vun,Shinohara:2021xry}, and to various other matter models~\cite{Yazadjiev:2015hda,Rogatko:2016mho}.

A natural open question is whether the constancy of $N$ on the photon surface is essential, or whether it can be relaxed in the uniqueness theorem. As a first step toward answering this question, Yoshino~\cite{Yoshino:2016kgi} adopted a perturbative approach and studied static first-order perturbations of the Schwarzschild spacetime. The photon surface in the distorted spacetime is assumed to be located at
$r = f(\theta, \phi)$, expanded as $f = 3M + \epsilon f^{(1)} + \cdots$. Imposing the umbilical condition~\eqref{eq:umbilicaleq:umbilical} on this surface, one finds that the first-order displacement $f^{(1)}$ must vanish, and the metric perturbation must satisfy a special constraint at $r = 3M$.
For any $\ell \geq 2$ mode of the even-parity perturbation, this constraint cannot be satisfied if the exterior region is vacuum, because the regular solution of the metric perturbation does not fulfill the required condition. Consequently, the \textit{perturbative uniqueness theorem} holds: there is no solution branch of a spacetime with a distorted photon surface that regularly connects to the Schwarzschild solution, provided the exterior spacetime is vacuum and asymptotically flat.

On the basis of this perturbative result, Yoshino proposed the following conjecture:
\begin{quote}
    \textit{If an asymptotically flat, vacuum spacetime possesses a static photon
    surface, then the spacetime is isometric to the Schwarzschild spacetime.}
\end{quote}
This conjecture suggests that the assumption of constant lapse function may be removable in a full non-perturbative uniqueness theorem, a direction that remains an important open problem.

It is worth noting that if matter is present outside $r = 3M$, the perturbative analysis allows for the existence of a distorted photon surface in principle. However, this requires a precise fine-tuning of the ratio $\alpha_\ell / \beta_\ell$ between the two linearly independent perturbative solutions, meaning that matter must be distributed in a highly special manner. Whether such configurations can be realized in a fully nonlinear setting, and whether they are physically relevant in an astrophysical context, remain open questions.

Although the uniqueness theorem discussed in this subsection is not directly invoked in the subsequent calculations, it provides the motivation for choosing the Schwarzschild spacetime as the background geometry. Under the assumptions of the uniqueness theorem, the spacetime outside the relevant photon surface is uniquely determined. We therefore take the Schwarzschild spacetime as the background and consider higher-curvature EFT corrections to this geometry. This provides a natural framework for investigating how such corrections modify the photon-sphere properties, gravitational lensing, and the QNMs of a test scalar field.

\section{Gravitational Effective Field Theory}
\label{sec.EFT corrections}
 
In this work, we investigate classical corrections to QNMs and gravitational lensing within the framework of effective field theory, with the aim of exploring potential constraints on theories beyond general relativity.

We consider four-dimensional vacuum spacetimes, for which the Einstein equations reduce to
\begin{align}
  R_{\mu\nu} = 0,
\end{align}
implying $R = 0$. In this case, the Riemann tensor coincides with the Weyl tensor,
\begin{align}
  R_{\mu\nu\rho\sigma} = C_{\mu\nu\rho\sigma}.
\end{align}
Therefore, higher-curvature operators can be constructed solely from the Riemann tensor.

In four dimensions, curvature-squared invariants reduce, up to tensor identities~\cite{Fulling:1992vm}, to the Gauss-Bonnet combination, which is topological and does not affect the bulk equations of motion. As a result, the leading nontrivial corrections arise at cubic order in the curvature.

The effective action is organized as a derivative expansion including all independent diffeomorphism-invariant operators, with higher-derivative terms suppressed by the cutoff scale. At cubic order, the independent operator is given by
\begin{align}
R^{\rho\sigma}_{\ \ \mu\nu}R^{\alpha\beta}_{\ \ \rho\sigma}R^{\mu\nu}_{\ \ \alpha\beta},
\end{align}
while at quartic order there are two independent invariants,
\begin{align}
(R_{\mu\nu\rho\sigma}R^{\mu\nu\rho\sigma})^2, \quad (R_{\mu\nu\rho\sigma}\tilde{R}^{\mu\nu\rho\sigma})^2.
\end{align}

Although cubic curvature terms are often expected to provide the leading corrections, it has been argued in Ref.~\cite{Camanho:2014apa,Endlich:2017tqa} that this need not always be the case. Motivated by this observation, we include quartic curvature invariants on the same footing and analyze their effects alongside the cubic terms. Accordingly, we consider the following effective action in four-dimensional spacetime:
\begin{align}
S=\frac{1}{16\pi G}\int d^4x\sqrt{-g}\left(R+\gamma R^{\rho\sigma}_{\ \ \mu\nu}R^{\alpha\beta}_{\ \ \rho\sigma}R^{\mu\nu}_{\ \ \alpha\beta}+\eta (R_{\mu\nu\rho\sigma}R^{\mu\nu\rho\sigma})^2+\tilde{\eta} (R_{\mu\nu\rho\sigma}\tilde{R}^{\mu\nu\rho\sigma})^2\right),
\end{align}
where the dual tensor is defined as
\begin{align}
\tilde{R}_{\mu\nu\rho\sigma}=\frac{1}{2}\epsilon_{\mu\nu}^{\ \ \alpha\beta}R_{\alpha\beta\rho\sigma}.
\end{align}
The parity-odd cubic invariant and the mixed quartic invariant, given by the product of the Kretschmann scalar and the Chern--Pontryagin density, are not considered here, as neither contributes to the static, spherically symmetric background at linear order in the EFT expansion. Consequently, the corresponding parity-odd coupling cannot be constrained by the observables considered in the present analysis.
 
We classify the local scalar curvature invariants relevant to the static, spherically symmetric background considered in this work, focusing on those that contribute at linear order in the EFT perturbative expansion. Up to the use of the metric and integration by parts, the independent scalar invariants that are cubic in the Riemann tensor in four dimensions can be organized as follows:
\begin{align}
  \mathcal{I}_1 &= R_{\mu\nu}{}^{\rho\sigma}
                   R_{\rho\sigma}{}^{\lambda\kappa}
                   R_{\lambda\kappa}{}^{\mu\nu}, \label{eq:I1} \\
  \mathcal{I}_2 &= R_{\mu}{}^{\nu}{}_{\rho}{}^{\sigma}
                   R_{\nu}{}^{\lambda}{}_{\sigma}{}^{\kappa}
                   R_{\lambda}{}^{\mu}{}_{\kappa}{}^{\rho}. \label{eq:I2}
\end{align}
These correspond to two inequivalent ways of fully contracting three copies of the Riemann tensor, distinguished by their index contraction structures. All other cubic invariants involving the Ricci tensor or Ricci scalar vanish identically in vacuum and are therefore omitted.
 
The quadratic Lovelock (Gauss-Bonnet) scalar
\begin{align}
  \mathcal{G} = R^2 - 4R_{\mu\nu}R^{\mu\nu} + R_{\mu\nu\rho\sigma}R^{\mu\nu\rho\sigma}
\end{align}
is a total derivative in four dimensions and does not contribute to the
equations of motion.
  
The cubic Lovelock scalar is defined by
\begin{align}
  \mathcal{L}_3 = \delta^{\mu_1\mu_2\mu_3\mu_4\mu_5\mu_6}
                  _{\nu_1\nu_2\nu_3\nu_4\nu_5\nu_6}
                  R^{\nu_1\nu_2}{}_{\mu_1\mu_2}
                  R^{\nu_3\nu_4}{}_{\mu_3\mu_4}
                  R^{\nu_5\nu_6}{}_{\mu_5\mu_6},
\end{align}
where $\delta^{\mu_1\cdots\mu_6}_{\nu_1\cdots\nu_6}$ is the generalised
Kronecker delta.

Expanding in terms of the Riemann, Ricci, and Ricci scalar gives
\begin{align}
  \mathcal{L}_3 = \; &R^3
    - 12\,R\,R_{\mu\nu}R^{\mu\nu}
    + 16\,R_{\mu\nu}R^{\nu\rho}R_\rho{}^\mu
    + 24\,R_{\mu\nu}R^{\mu\rho\nu\sigma}R_{\rho\sigma} \notag\\
    &+ 3\,R\,R_{\mu\nu\rho\sigma}R^{\mu\nu\rho\sigma}
    - 24\,R_{\mu\nu\rho\sigma}R^{\mu\nu\rho}{}_\lambda R^{\sigma\lambda}
    + 4\,R_{\mu\nu\rho\sigma}R^{\mu\nu\lambda\kappa}R^\rho{}_\lambda{}^\sigma{}_\kappa \notag\\
    &- 8\,R_{\mu\rho\nu\sigma}R^{\mu\lambda\nu\kappa}R^\rho{}_\lambda{}^\sigma{}_\kappa.
    \label{eq:L3}
\end{align}
 
In four spacetime dimensions, $\mathcal{L}_3 \equiv 0$ identically. This follows because the generalised Kronecker delta
$\delta^{\mu_1\cdots\mu_6}_{\nu_1\cdots\nu_6}$ requires complete antisymmetrisation over six distinct indices, whereas four-dimensional spacetime has only four independent directions; any such antisymmetrisation therefore vanishes identically.
 
Imposing the vacuum condition $R_{\mu\nu} = 0$ (and hence $R = 0$) on \eqref{eq:L3}, every term containing $R_{\mu\nu}$ or $R$ drops out, leaving
\begin{align}
  0 = \mathcal{L}_3\big|_{R_{\mu\nu}=0}= 4\,\mathcal{I}_1 - 8\,\mathcal{I}_2.
\end{align}
 
This yields the algebraic relation
\begin{align}
  \mathcal{I}_2 = \frac{1}{2}\,\mathcal{I}_1. \label{eq:relation}
\end{align}
 
In a four-dimensional vacuum spacetime ($R_{\mu\nu} = 0$), the two independent cubic Riemann invariants $\mathcal{I}_1$ and $\mathcal{I}_2$ satisfy \eqref{eq:relation}, so there is only \emph{one} algebraically independent cubic scalar invariant, which may be taken to be
\begin{align}
    \mathcal{I}_1 = R_{\mu\nu}{}^{\rho\sigma}R_{\rho\sigma}{}^{\lambda\kappa}R_{\lambda\kappa}{}^{\mu\nu}
                  = C_{\mu\nu}{}^{\rho\sigma}C_{\rho\sigma}{}^{\lambda\kappa}C_{\lambda\kappa}{}^{\mu\nu}.
\end{align}
 
In four-dimensional vacuum, a natural and convenient pair of independent quartic scalar invariants is
\begin{align}
  \mathcal{J}_1 &= \left(R_{\mu\nu\rho\sigma}R^{\mu\nu\rho\sigma}\right)^2, \label{eq:J1}\\
  \mathcal{J}_{\mathrm{H}} &= \left(R_{\mu\nu\rho\sigma}\tilde{R}^{\mu\nu\rho\sigma}\right)^2.
  \label{eq:JH}
\end{align}
In vacuum ($R_{\mu\nu\rho\sigma} = C_{\mu\nu\rho\sigma}$) these become
\begin{align}
  \mathcal{J}_1 &= \left(C_{\mu\nu\rho\sigma}C^{\mu\nu\rho\sigma}\right)^2, \\
  \mathcal{J}_{\mathrm{H}} &= \left(C_{\mu\nu\rho\sigma}\tilde{C}^{\mu\nu\rho\sigma}\right)^2,
\end{align}
 where the Hodge dual of the Weyl tensor is defined as
\begin{equation}
  \tilde{C}^{\mu\nu\rho\sigma}
  = \frac{1}{2}\epsilon^{\mu\nu\lambda\kappa}\,C_{\lambda\kappa}{}^{\rho\sigma}.
\end{equation}

The quantity $R_{\mu\nu\rho\sigma}\tilde{R}^{\mu\nu\rho\sigma}$ is the Chern-Pontryagin density, which equals a total derivative in four dimensions:
\begin{align}
  R_{\mu\nu\rho\sigma}\tilde{R}^{\mu\nu\rho\sigma} = \nabla_\mu K^\mu.
\end{align}
This reflects the topological nature of the parity-odd sector at quadratic order. Although it is itself a total derivative (and hence does not contribute to the action), its \emph{square} $\mathcal{J}_{\mathrm{H}}$ is a genuine
local scalar invariant that is \emph{not} a total derivative and carries independent physical information.
 
The independence of $\mathcal{J}_1$ and $\mathcal{J}_{\mathrm{H}}$ can be seen from the self-dual/anti-self-dual decomposition of the Weyl tensor,
\begin{equation}
  C_{\mu\nu\rho\sigma} = C^+_{\mu\nu\rho\sigma} + C^-_{\mu\nu\rho\sigma},
  \qquad \tilde{C}^\pm_{\mu\nu\rho\sigma} = \pm C^\pm_{\mu\nu\rho\sigma},
\end{equation}
which gives
\begin{align}
  C_{\mu\nu\rho\sigma}C^{\mu\nu\rho\sigma}
    &= C^+_{\mu\nu\rho\sigma}C^{+\mu\nu\rho\sigma}
     + C^-_{\mu\nu\rho\sigma}C^{-\mu\nu\rho\sigma}
     \equiv \mathcal{A} + \mathcal{B}, \\
  C_{\mu\nu\rho\sigma}\tilde{C}^{\mu\nu\rho\sigma}
    &= C^+_{\mu\nu\rho\sigma}C^{+\mu\nu\rho\sigma}
     - C^-_{\mu\nu\rho\sigma}C^{-\mu\nu\rho\sigma}
     \equiv \mathcal{A} - \mathcal{B}.
\end{align}
Hence
\begin{equation}
  \mathcal{J}_1 = (\mathcal{A}+\mathcal{B})^2, \qquad
  \mathcal{J}_{\mathrm{H}} = (\mathcal{A}-\mathcal{B})^2.
\end{equation}
Since we consider a parity-even effective theory in this work, $(\mathcal{J}1,\mathcal{J}{\mathrm{H}})$ can be chosen as a basis for the quartic curvature invariants relevant to our analysis. This choice is further supported by the quartic Lovelock identity, to which we now turn.
 
The quartic Lovelock scalar
\begin{equation}
  \mathcal{L}_4 = \delta^{\mu_1\cdots\mu_8}_{\nu_1\cdots\nu_8}
                  R^{\nu_1\nu_2}{}_{\mu_1\mu_2}
                  R^{\nu_3\nu_4}{}_{\mu_3\mu_4}
                  R^{\nu_5\nu_6}{}_{\mu_5\mu_6}
                  R^{\nu_7\nu_8}{}_{\mu_7\mu_8}
\end{equation}
vanishes identically in four dimensions.

\begin{table}[htbp]
\centering
\caption{Independent cubic curvature invariants in vacuum ($R_{\mu\nu}=0$), where the Riemann tensor coincides with the Weyl tensor. The invariant $\mathcal{I}_2$ is related to $\mathcal{I}_1$ by the Lovelock identity, so $\mathcal{I}_1$ is the only independent cubic invariant.}
\label{tab:cubic}
\renewcommand{\arraystretch}{1.5}
\begin{tabular}{lll}
\hline
Invariant & Value in vacuum & Independent? \\
\hline
$\mathcal{I}_1 = R_{\mu\nu}{}^{\rho\sigma}R_{\rho\sigma}{}^{\lambda\kappa}R_{\lambda\kappa}{}^{\mu\nu}$
  & $= C_{\mu\nu}{}^{\rho\sigma}C_{\rho\sigma}{}^{\lambda\kappa}C_{\lambda\kappa}{}^{\mu\nu}$
  & \textbf{Yes (unique)} \\[4pt]
$\mathcal{I}_2 = R_{\mu}{}^{\nu}{}_{\rho}{}^{\sigma}R_{\nu}{}^{\lambda}{}_{\sigma}{}^{\kappa}R_{\lambda}{}^{\mu}{}_{\kappa}{}^{\rho}$
  & $= \tfrac{1}{2}\,\mathcal{I}_1$
  & No (dependent via Lovelock identity) \\
\hline
\end{tabular}
\end{table}

\begin{table}[htbp]
\centering
\caption{Independent even-parity quartic curvature invariants in vacuum. Here $\tilde{R}^{\mu\nu\rho\sigma}$ and $\tilde{C}^{\mu\nu\rho\sigma}$ denote the duals of the Riemann and Weyl tensors. Both $\mathcal{J}_1$ and $\mathcal{J}_{\mathrm{H}}$ are parity-even. All other even-parity quartic contractions are expressible in terms of $\mathcal{J}_1$ and $\mathcal{J}_{\mathrm{H}}$.}
\label{tab:quartic}
\renewcommand{\arraystretch}{1.5}
\begin{tabular}{lll}
\hline
Invariant & Value in vacuum & Independent? \\
\hline
$\mathcal{J}_1 = \bigl(R_{\mu\nu\rho\sigma}R^{\mu\nu\rho\sigma}\bigr)^2$
  & $= \bigl(C_{\mu\nu\rho\sigma}C^{\mu\nu\rho\sigma}\bigr)^2$
  & \textbf{Yes} \\[4pt]
$\mathcal{J}_{\mathrm{H}} = \bigl(R_{\mu\nu\rho\sigma}\tilde{R}^{\mu\nu\rho\sigma}\bigr)^2$
  & $= \bigl(C_{\mu\nu\rho\sigma}\tilde{C}^{\mu\nu\rho\sigma}\bigr)^2$
  & \textbf{Yes} \\[4pt]
Other quartic contractions
  & expressible via $\mathcal{J}_1,\mathcal{J}_{\mathrm{H}}$
  & No \\
\hline
\end{tabular}
\end{table}

\noindent
For the static, spherically symmetric background considered in this work, the independent even-parity local scalar curvature invariants at cubic and quartic order in the curvature can be organized as
\begin{itemize}
\item cubic: $\mathcal{I}_1$,
\item quartic: $\mathcal{J}_1$ and $\mathcal{J}_{\mathrm{H}}$.
\end{itemize}
The pair $(\mathcal{J}_1,\mathcal{J}_{\mathrm{H}})$ can equivalently be expressed in terms of the norms of the self-dual and anti-self-dual components of the Weyl tensor, $(\mathcal{A},\mathcal{B})$, via
$\mathcal{A} = \tfrac{1}{2}(\sqrt{\mathcal{J}_1}+\sqrt{\mathcal{J}_{\mathrm{H}}})$
and
$\mathcal{B} = \tfrac{1}{2}(\sqrt{\mathcal{J}_1}-\sqrt{\mathcal{J}_{\mathrm{H}}})$.

In four-dimensional vacuum spacetime, the square of the Riemann tensor coincides with the Gauss--Bonnet density. Since the Gauss-Bonnet term is topological and does not contribute to the equations of motion, curvature-squared terms do not contribute to the static, spherically symmetric background at the order considered here. We therefore focus on the curvature invariants that contribute to the static, spherically symmetric background at linear order in the EFT expansion. The parity-odd cubic invariant and the mixed quartic invariant given by the product of the Kretschmann scalar and the Chern--Pontryagin density are not included, since they do not contribute to the background geometry at this order. Consequently, the corresponding couplings cannot be constrained by the observables considered in this work.

\section{Photon surface with higher curvature corrections in the Schwarzschild}
\label{sec.photon surface EFT}

In this section, we determine how higher-curvature corrections modify the Schwarzschild spacetime by solving the field equations perturbatively to first order. In particular, we compute the resulting shift in the location of the photon surface.

The gravitational theory under consideration is diffeomorphism invariant, and in general the coordinates serve merely as labels of spacetime points without direct physical significance. To assign a physical meaning to the radial coordinate, we fix the gauge by identifying it with the areal radius, such that surfaces of constant radial coordinate have area $4\pi r^2$. In this gauge, the value of the radial coordinate acquires an invariant geometrical interpretation, allowing us to unambiguously characterize the displacement of the photon surface.

As discussed in the previous section, the form of the effective action is fixed by the classification of independent higher-curvature invariants in four-dimensional vacuum spacetimes. In particular, curvature-squared terms do not contribute to the equations of motion, and the leading nontrivial corrections arise at cubic order in the Riemann tensor, with quartic terms included on the same footing. Then we consider the action;
\begin{equation}
S=\frac{1}{16\pi G}\int d^4x\sqrt{-g}(R+\frac{1}{2}\gamma\mathcal{I}_1+\frac{1}{2}\eta\mathcal{J}1+\frac{1}{2}\tilde{\eta} \mathcal{J}_{\mathrm H}).
\end{equation}
We use the invariants $\mathcal{I}_1$, $\mathcal{J}1$ and $\mathcal{J}_{\mathrm H}$ defined in Sec.~\ref{sec.EFT corrections}. 

We now derive the equations of motion from the above action.
\begin{align}
R_{\mu\nu}-\frac{1}{2}g_{\mu\nu}R=\gamma T_{\mu\nu}^{{\rm cubic}}+\eta T_{\mu\nu}^{{\rm quartic}}+\tilde{\eta}\tilde{T}_{\mu\nu}^{{\rm quartic}},
\end{align}
where $T^{\rm cubic}_{\mu\nu}$, $T^{\rm quartic}_{\mu\nu}$ and $\tilde T^{\rm quartic}_{\mu\nu}$ are defined as the energy-momentum tensors obtained from the metric variations of the $\gamma\,\mathcal{I}_1$, $\eta\,\mathcal{J}_1$ and $\tilde\eta\,\mathcal{J}_{\mathrm H}$ terms, respectively, with the couplings $\gamma$, $\eta$ and $\tilde\eta$ factored out:
\begin{align}
T_{\mu\nu}^{{\rm cubic}}&=\frac{3}{2}R_{\mu}^{\ \rho\sigma\lambda}R_{\sigma\lambda}^{\ \ \ \alpha\beta}R_{\alpha\beta\rho\nu}+\frac{1}{4}g_{\mu\nu}R_{\alpha\beta}^{\ \ \ \rho\sigma}R_{\rho\sigma}^{\ \ \ \lambda\gamma}R_{\lambda\gamma}^{\ \ \alpha\beta}-3\nabla^\rho\nabla^\sigma(R_{\mu\rho\alpha\beta}R_{\nu\sigma}^{\ \ \alpha\beta}),\\
T_{\mu\nu}^{{\rm quartic}}&=-4R_{\mu\rho\nu\sigma}\nabla^\rho\nabla^\sigma(R_{\mu\nu\rho\sigma}R^{\mu\nu\rho\sigma})-\frac{1}{4}g_{\mu\nu}(R_{\mu\nu\rho\sigma}R^{\mu\nu\rho\sigma})^2,\\
\tilde{T}_{\mu\nu}^{{\rm quartic}}&=-4\tilde{R}_{\mu\rho\nu\sigma}\nabla^\rho\nabla^\sigma(R_{\mu\nu\rho\sigma}\tilde{R}^{\mu\nu\rho\sigma})-\frac{1}{4}g_{\mu\nu}(R_{\mu\nu\rho\sigma}\tilde{R}^{\mu\nu\rho\sigma})^2.
\end{align}
We note that $\mathcal{J}_{\mathrm H}$ does not contribute to the static, spherically symmetric background at first order in the perturbative expansion.

We study static first-order perturbations of a Schwarzschild spacetime. By virtue of the uniqueness theorem, any static, asymptotically flat vacuum spacetime is given by the Schwarzschild solution \cite{Israel:1967wq}, which we therefore take as the background. The background metric is given in the coordinates $(t,r,\theta,\phi)$ as follows:
\begin{equation}
ds^2=g_{\mu\nu}dx^{\mu}dx^{\nu}=-\left(1-\frac{2GM}{r}\right)dt^2+\frac{dr^2}{\left(1-\frac{2GM}{r}\right)}+r^2(d\theta^2+\sin^2\theta d\phi^2).
\end{equation}
We consider static perturbations of the Schwarzschild spacetime and introduce a small expansion parameter $\epsilon$. In the present work, we consider higher-curvature terms that are suppressed by a small parameter $\epsilon$, and we treat them as corrections of order $\mathcal{O}(\epsilon)$. The distorted metric is written as
\begin{equation}
ds^2
=
- e^{2\nu} dt^2
+ e^{2\mu} dr^2
+ r^2
\left(
d\theta^2 + \sin^2\theta\, d\phi^2
\right),
\end{equation}
where the metric functions are expanded as
\begin{align}
\nu &= \nu^{(0)} + \epsilon\, \nu^{(1)} + \cdots, \\
\mu &= \mu^{(0)} + \epsilon\, \mu^{(1)} + \cdots, \\
e^{2\nu^{(0)}}&=e^{-2\mu^{(0)}}=1-\frac{2GM}{r}.
\end{align}
The couplings $\gamma$ and $\eta$ are dimensionful, with $[\gamma]=(\text{length})^4$ and $[\eta]=(\text{length})^6$. For the Schwarzschild background, the only length scale is the horizon radius $r_s = 2GM$, so we normalize them by appropriate powers of $r_s$ and define the dimensionless parameters
\begin{equation}
  \epsilon_\gamma \equiv \frac{\gamma}{r_s^{4}}, \qquad
  \epsilon_\eta \equiv \frac{\eta}{r_s^{6}} .
  \label{eq:eps}
\end{equation}
Writing $\gamma\sim\Lambda^{-4}$ and $\eta\sim\Lambda^{-6}$ in terms of the EFT cutoff $\Lambda$, both are controlled by a single small parameter,
\begin{equation}
  \epsilon \equiv \frac{1}{\Lambda r_s}, \qquad
  \epsilon_\gamma \sim \epsilon^{4}, \qquad
  \epsilon_\eta \sim \epsilon^{6}.
  \label{eq:eps_single}
\end{equation}
Each higher-curvature term is thus suppressed relative to the Einstein--Hilbert term by the corresponding power of $\epsilon$, multiplied by a function of $r/r_s$.

The perturbative expansion is valid for
\begin{equation}
  \epsilon \ll 1
  \quad\Longleftrightarrow\quad
  \Lambda r_s \gg 1,
  \label{eq:validity}
\end{equation}
that is, when the curvature scale of the black hole is well below the cutoff. In this regime $|\epsilon_\gamma|\ll1$ and $|\epsilon_\eta|\ll1$, and all results are computed to linear order in $\epsilon_\gamma$ and $\epsilon_\eta$. The curvature invariants of the Schwarzschild background scale as $\mathcal{R}\sim r_s/r^3$, so the corrections are largest at the horizon, and Eq.~\eqref{eq:validity} ensures that they remain small everywhere outside it. Terms of higher order in the curvature are further suppressed by additional powers of $\epsilon$ and can be consistently neglected. In particular, $\epsilon_\gamma^2\sim\epsilon^{8}$ is parametrically smaller than $\epsilon_\eta\sim\epsilon^{6}$, so keeping only linear terms is consistent.

The first-order perturbations are expanded in spherical harmonics,
\begin{align}
\nu^{(1)} &=  \sum_{\ell,m} K^{(1)}_{\ell m}(r)\,
Y_{\ell m}(\theta,\phi), \\
\mu^{(1)} &= \sum_{\ell,m} H^{(1)}_{\ell m}(r)\,
Y_{\ell m}(\theta,\phi),
\end{align}
where $Y_{\ell m}(\theta,\phi)$ denote the spherical harmonics with
$\ell \ge 0$ and $-\ell \le m \le \ell$.

At linear order, perturbations with different $(\ell,m)$ decouple, so that each mode can be analyzed independently. In the following, we therefore focus on a single mode and suppress the indices $(\ell,m)$, denoting the radial functions simply by $U(r):=-K^{(1)}(r)$ and $T(r):=H^{(1)}(r)$.

In this work, we consider the Schwarzschild spacetime as the background solution and incorporate higher-curvature corrections within an effective field theory (EFT) framework. Treating these higher-curvature terms as a source, we solve the Einstein equations perturbatively to obtain the corrected solution. Owing to the spherical symmetry of the Schwarzschild background, the induced source term is also spherically symmetric. As a result, the perturbative solution preserves spherical symmetry, and only the $\ell=0$, $m=0$ mode is present.

This can also be understood from the uniqueness of static, asymptotically flat vacuum black hole solutions in four dimensions, which implies that the Schwarzschild spacetime is the only such solution \cite{Israel:1967wq}. Consequently, no nontrivial higher-multipole perturbations arise in this setup. The radial equations for the first-order perturbation variables are then given by
\begin{align}
&T(r)\, r^{10}+ 24M^{2}\left(24 \eta M (134M - 64 r)+ \gamma (98M - 45 r)\, r^{3}\right)+ r^{10} (r - 2M)\, T'(r)=0,\\
&T(r)\, r^{10}+ 24M^{2}\left(24 \eta M (22M - 8 r)+ \gamma r^{3} (-10M + 9 r)\right)+ r^{10} (r - 2M)\, U'(r)=0,\\
&\frac{48 M^{2}\left(24 \eta M (82 M - 36 r)+ \gamma r^{3} (-62 M + 27 r)\right)}{r^{10}}- (r+M)\, U'(r)-(r-M)\, T'(r)-(r-2M)\, r\, U''(r)=0.
\end{align}

Solving these equations, we obtain the perturbative solution as follows:
\begin{align}
U(r)&=\frac{1}{r^{9} (r - 2M)^{3/2}}\Big(64\eta M^{3} (22M - 16 r) \sqrt{r - 2M}- 2 (2M - r) r^{9} \nonumber\\
&\ \ \ \ - r^{3} \Big(40\gamma M^{3} \sqrt{r - 2M}+ r^{6} \sqrt{r -2 M}\, C_{1}\Big)\Big)+ C_{2},\\
T(r)&=\frac{24M^{2}}{r-2M}\left(\frac{1072 \eta M^{2}}{3 r^{9}}- \frac{192 \eta M}{r^{8}}+ \frac{49 \gamma M}{3 r^{6}}-\frac{9 \gamma}{r^{5}}\right)
+\frac{C_{1}}{r-2M}.
\end{align}
At first sight, the perturbative solution seems to possess a divergence at $r=2M$. However, evaluating the Kretschmann scalar reveals that the geometry is in fact regular at this location.
\begin{align}
K=\frac{48M^2}{r^6}+\frac{16 M}{r^{15}}\left(768 \eta M^{3} (212 M - 99 r)+ 36\gamma M^{2} (50M - 24 r) r^{3}- 3 r^{9} C_{1}\right),
\end{align}
At $\mathcal{O}(\epsilon)$, the perturbative solution generically contains a correction proportional to $C_1/r$, which can be interpreted as a shift of the black hole mass. Since such a term has the same radial dependence as the Schwarzschild contribution, it does not introduce a genuinely new physical effect but instead corresponds to a redefinition of the mass parameter.

In asymptotically flat spacetimes, the physically relevant mass is given by the ADM mass, which is determined by the coefficient of the $1/r$ term in the asymptotic expansion of the metric. Therefore, it is natural to absorb this correction into a redefined mass parameter,
\begin{equation}
M_{\rm phys} = M + \epsilon\,\delta M,
\end{equation}
so that the metric retains the standard Schwarzschild form at large distances.

In the following analysis, we express all quantities in terms of the physical mass $M_{\rm phys}$ and focus on genuine higher-order corrections, such as those proportional to $1/r^2$ or higher powers. This procedure allows us to isolate the true effects of the underlying theory, eliminating spurious contributions associated with a mere rescaling of the mass. Imposing the condition $g_{tt} \to 1$ as $r \to \infty$ and the regularity at $r=2M$ lead to $ C_1= C_2  = 0$, and the result is therefore given as follows.
\begin{align}
U(r)&=\frac{1}{r^{9} (r - 2M)}\Big(64\eta M^{3} (22M - 16 r)- 40\gamma M^{3}r^3\Bigr),\\
T(r)&=\frac{24M^{2}}{r-2M}\left(\frac{1072 \eta M^{2}}{3 r^{9}}- \frac{192 \eta M}{r^{8}}+ \frac{49 \gamma M}{3 r^{6}}- \frac{9 \gamma}{r^{5}}\right),
\end{align}

The zeroth order of the extrinsic curvature on  $r=const.$ surface is
\begin{align}
K_{\mu\nu}^{(0)}dx^{\mu}dx^{\nu}=-\frac{M}{r^2}\left(1-\frac{2M}{r}\right)^{\frac{1}{2}}dt^2+r\left(1-\frac{2M}{r}\right)^{\frac{1}{2}}(d\theta^2+\sin^2\theta d\phi^2),
\end{align}
The zeroth order of the induced metric on  $r=const.$ surface is
\begin{align}
h_{\mu\nu}^{(0)}dx^{\mu}dx^{\nu}=-\left(1-\frac{2M}{r}\right)dt^2+r^2(d\theta^2+\sin^2\theta d\phi^2),
\end{align}
The third condition~\eqref{eq:umbilicaleq:umbilical} of photon surface in the zeroth order is satisfied on $r=3M$.

For the next step, we consider the perturbation up to first order of the photon surface. Accordingly, the extrinsic curvature up to first order is given by
\begin{align}
K_{\mu\nu}dx^{\mu}dx^{\nu}&=K_{tt}dt^2+K_{\theta\theta}(d\theta^2+\sin^2\theta d\phi^2),\\
K_{tt}&:=-\frac{M}{r^2}\left(1-\frac{2M}{r}\right)^{\frac{1}{2}}+\frac{2M^3(9184\eta M^2-9280\eta Mr+2304\eta r^2-42\gamma Mr^3+16\gamma r^4)}{r^{12}}\frac{1}{\left(1-\frac{2M}{r}\right)},\\
K_{\theta\theta}&:=r\left(1-\frac{2M}{r}\right)^{\frac{1}{2}}+\frac{2M^2(-2144\eta M^2+1152\eta Mr-98\gamma Mr^3+54\gamma r^4)}{r^9}\frac{1}{\left(1-\frac{2M}{r}\right)}.
\end{align}
In the same manner, the induced metric up to first order is given by
\begin{align}
h_{\mu\nu}dx^{\mu}dx^{\nu}&=h_{tt}dt^2+h_{\theta\theta}(d\theta^2+\sin^2\theta d\phi^2),\\
h_{tt}&=-\left(1-\frac{2M}{r}\right)+\frac{8M^3(176\eta M-128\eta r-5\gamma r^3)}{r^{10}},\\
h_{\theta\theta}&=r^2.
\end{align}
The third condition~\eqref{eq:umbilicaleq:umbilical} for the photon surface, up to first order, is satisfied at the following value of $r$:
\begin{align}
r=3M-\frac{4(704\eta+405\gamma M^2)}{6561M^5}.
\end{align}
\begin{align}
\frac{h_{\theta\theta}}{h_{tt}}=-27M^2+\frac{8(208\eta+135\gamma M^2)}{729M^4},
\end{align}
We use the areal radius in this analysis. Since the area is a gauge-invariant quantity, the present result is independent of the choice of gauge.

\subsection{Impact Parameter and Photon Surface}

In this subsection, we briefly review the definition of the impact parameter and its relation to the photon surface. Owing to the spherical symmetry of the spacetime, the motion of the photon can be restricted without loss of generality to the equatorial plane, $\theta=\pi/2$. We therefore set $\theta=\pi/2$ and $\dot{\theta}=0$ in the following analysis.

We consider a static, spherically symmetric spacetime with the metric
\begin{equation}
ds^2 = -f(r)\,dt^2 + g(r)dr^2 + r^2 d\Omega^2 \, .
\end{equation}

Due to time-translation and rotational symmetries, the motion of a test particle admits two conserved quantities, namely the energy $E$ and the angular momentum $L$, given by
\begin{align}
E = f(r)\,\dot{t}, 
\qquad 
L = r^2 \dot{\phi} \, .
\end{align}
For null geodesics, the condition \(ds^2 = 0\) leads to the radial equation
\begin{align}
\dot{r}^2 + U_{\mathrm{eff}}(r) = \frac{1}{f(r)g(r)}E^2 \, ,
\end{align}
where the effective potential is defined as
\begin{align}
U_{\mathrm{eff}}(r) = \frac{L^2}{r^2g(r)} \, .
\end{align}

The impact parameter $u$ is defined as the ratio of the conserved quantities,
\begin{align}
u = \frac{L}{E} \, .
\end{align}
In terms of $b$, the radial equation can be rewritten as
\begin{align}
\dot{r}^2 = \frac{E^2}{f(r)g(r)} \left( 1 - \frac{u^2 f(r)}{r^2} \right) \, .
\end{align}

Circular null orbits correspond to the conditions
\begin{align}
\dot{r} = 0, 
\qquad 
\frac{dU_{\mathrm{eff}}}{dr} = 0 \, .
\end{align}
The first condition implies
\begin{align}
u^2 = \frac{r^2}{f(r)} \, .
\end{align}
Evaluating this expression at the radius of the photon surface $r = r_{ph}$, we obtain the critical impact parameter
\begin{align}
u_c^2 = \frac{r_{ph}^2}{f(r_{ph})} \, .
\end{align}

In the case of the Schwarzschild spacetime, where $f(r) =1/g(r)= 1 - 2M/r$, the photon surface is located at $r_{ph} = 3M$, leading to
\begin{align}
u_c = 3\sqrt{3}\, M \, .
\end{align}

The critical impact parameter \(u_c\) determines the boundary between capture and scattering of null geodesics and is directly related to the size of the black hole shadow. In the presence of higher-curvature corrections, both the metric function \(f(r)\) and the photon surface radius \(r_{ph}\) receive perturbative modifications, which in turn lead to corrections to \(u_c\).

\begin{align}
u_c=3\sqrt{3}M-\frac{4(208\eta+135\gamma M^2)}{2187\sqrt{3}M^5}.
\end{align}

This critical impact parameter characterizes the boundary between capture and scattering of null geodesics, and thus plays a central role in determining the strong gravitational lensing observables.

\section{Eikonal Quasinormal Modes in Effective Field Theory}
\label{sec:qnm}

Observables may provide a means to place constraints on effective field theories of gravity. In this section, we focus on quasinormal mode (QNM) frequencies as one such observable and compute the contributions of higher-curvature corrections within the eikonal approximation.

The metric of a general static spherically symmetric spacetime is given by
\begin{equation}
ds^2 = - f(r)\, dt^2 + g(r)\, dr^2 + r^2 (d\theta^2 + \sin^2\theta\, d\phi^2).
\end{equation}

We consider a massless scalar field $\Phi$, minimally coupled to gravity, as a test field propagating on this fixed background; its backreaction on the geometry is neglected. Our QNM analysis is thus restricted to the test-field limit and does not describe gravitational (tensor) perturbations of the metric itself, whose evolution is additionally modified by the higher-curvature operators. The scalar field serves as a simplified probe of the QNM spectrum of the corrected background. In the eikonal limit, QNM frequencies are determined by the properties of null geodesics near the photon sphere, which is the regime of interest in this work. The dynamics of $\Phi$ is governed by the covariant d'Alembertian operator defined as
\begin{equation}
\square \Phi
= \frac{1}{\sqrt{-g}} \partial_\mu \left( \sqrt{-g}\, g^{\mu\nu} \partial_\nu \Phi \right).
\end{equation}

To evaluate this explicitly, we first compute the determinant of the metric, which reads
\begin{equation}
\sqrt{-g} = \sqrt{f(r)\, g(r)} \; r^2 \sin\theta.
\end{equation}

We also need the inverse metric components, which are given by
\begin{align}
g^{tt} &= -\frac{1}{f(r)}, \\
g^{rr} &= \frac{1}{g(r)}, \\
g^{\theta\theta} &= \frac{1}{r^2}, \\
g^{\phi\phi} &= \frac{1}{r^2 \sin^2\theta}.
\end{align}

Substituting these expressions into the definition of the d'Alembertian, we obtain
\begin{equation}
\square \Phi =
- \frac{1}{f(r)} \partial_t^2 \Phi
+ \frac{1}{\sqrt{f g}\, r^2} \partial_r \left( \sqrt{\frac{f}{g}}\, r^2 \partial_r \Phi \right)
+ \frac{1}{r^2} \Delta_{S^2} \Phi,
\end{equation}
where $\Delta_{S^2}$ denotes the Laplacian on the unit two-sphere, explicitly given by
\begin{equation}
\Delta_{S^2}
= \frac{1}{\sin\theta} \partial_\theta (\sin\theta \partial_\theta)
+ \frac{1}{\sin^2\theta} \partial_\phi^2.
\end{equation}

In order to analyze the wave equation, we separate variables by expanding the scalar field in spherical harmonics:
\begin{equation}
\Phi(t,r,\theta,\phi)
= e^{-i\omega t} \, Y_{\ell m}(\theta,\phi)\, \frac{\psi(r)}{r}.
\end{equation}

Using the eigenvalue equation for spherical harmonics,
\begin{equation}
\Delta_{S^2} Y_{\ell m}
= -\ell(\ell+1) Y_{\ell m},
\end{equation}
the angular dependence can be separated out, leaving a radial equation for $\psi(r)$.

It is convenient to introduce the tortoise coordinate $r_*$ defined by
\begin{equation}
\frac{dr_*}{dr} = \sqrt{\frac{g(r)}{f(r)}},
\end{equation}
which simplifies the radial part of the wave equation into a Schr\"odinger-like form.

In terms of $r_*$, the radial equation can be written as
\begin{equation}
\frac{d^2 \psi}{dr_*^2}
+ \left( \omega^2 - V_{\text{eff}}(r) \right)\psi = 0,
\end{equation}
where $V_{\text{eff}}(r)$ is an effective potential.

After some straightforward algebra, the effective potential is found to be
\begin{equation}
V_{\text{eff}}(r)
= f(r)\left[
\frac{\ell(\ell+1)}{r^2}
+ \frac{1}{r\sqrt{f(r)g(r)}} \frac{d}{dr}\left(\sqrt{\frac{f(r)}{g(r)}}\right)
\right].
\label{eff potential}
\end{equation}

In the next subsection, we compute the effective potential in the eikonal approximation for the Schwarzschild spacetime with higher-curvature corrections derived in the previous section, and determine the resulting quasinormal mode frequencies.

We next compute the QNMs in the eikonal approximation for the perturbed Schwarzschild spacetime derived in the previous section. The eikonal limit corresponds to the geometric optics regime, where the wavelength $\lambda_w$ is much smaller than the curvature scale $L_{\rm curv}$:
\begin{align}
\lambda_w \ll L_{\rm curv}.
\end{align}

The eikonal limit is defined by $\lambda \ll L$, where $\lambda$ denotes the wavelength of the perturbation and $L$ is the characteristic length scale of the background geometry. Since $\ell$ characterizes the angular scale of the perturbation, this condition corresponds to $\ell \gg 1$. In this limit, the second term in Eq.~(\ref{eff potential}) is suppressed by $O(\ell^{-1})$ relative to the leading term. Therefore, retaining only the leading contribution in the large-$\ell$ expansion, Eq.~(\ref{eff potential}) reduces to
\begin{equation}
\frac{d}{dr} \left( \frac{f(r)}{r^2} \right) = 0,
\end{equation}
which is precisely the condition for unstable circular null geodesics, defining the photon sphere. In a static, spherically symmetric spacetime, the photon sphere coincides with a photon surface only for the corresponding specific hypersurface $r={\rm const.}$, and the defining condition for this photon surface reduces to the above relation.

The potential has a single maximum (the peak of the potential barrier) at $r_*=r_*^{\rm max}$, which, in the eikonal limit, is located at the photon sphere. We define
\begin{align}
  V_0 &\equiv V\big(r_*^{\rm max}\big),\\
  V_0'' &\equiv \left.\frac{d^2 V}{dr_*^2}\right|_{r_*=r_*^{\rm max}} ,
\end{align}
that is, $V_0$ is the value of the effective potential at its peak, and $V_0''$ is its second derivative with respect to the tortoise coordinate $r_*$ evaluated at the peak. Since the peak is a maximum, $V_0''<0$. In the lowest-order WKB approximation~ \cite{Schutz:1985km,Iyer:1986np,Iyer:1986nq,Cardoso:2008bp,osti_4677534,Mashhoon:1985cya}, the QNM frequencies are then given by
\begin{align}
\omega_{\ell n}
&\simeq \omega_R - i\omega_I \\
&= \sqrt{V_0} - i\left(n+\frac{1}{2}\right)\sqrt{\frac{-V_0''}{2V_0}},
\end{align}
where $n=0,1,2,\dots$ is the overtone number, $\omega_R$ determines the oscillatory behavior of the perturbation, whereas $\omega_I$ determines the rate at which the amplitude decays with time. In particular, a larger positive value of $\omega_I$ corresponds to faster damping, while a smaller value corresponds to slower damping.

In this limit, the effective potential is dominated by the centrifugal term $\ell(\ell+1)/r^2$, which explains the universal behavior of different spin fields at high angular momentum. Subleading contributions encode spin-dependent corrections and become relevant only for small $\ell$. Consequently, the scalar-field analysis captures the universal high-frequency behavior of perturbations.

Expanding the potential around its maximum, we obtain
\begin{align}
V_0 &= \frac{\ell(\ell+1)}{27M^2}
+ \frac{8(208\eta+135\gamma M^2)\ell(\ell+1)}{531441M^8}, \\
V_0'' &= -\frac{2\ell(\ell+1)}{729M^4}
+ \frac{128(200\eta+27\gamma M^2)\ell(\ell+1)}{14348907M^{10}}.
\end{align}
To make the effect of the higher-curvature corrections on the QNM spectrum explicit, we write the eikonal QNM frequency as
\begin{align}
\omega_R
&=\omega_R^{\rm Schw}\left[1+\frac{540}{19683}\frac{\gamma}{M^4}+\frac{832}{19683}\frac{\eta}{M^6}\right],\\
\omega_I
&=
\omega_I^{\rm Schw}
\left[
1+\frac{1404}{19683}\frac{\gamma}{M^4}+\frac{7232}{19683}\frac{\eta}{M^6}\right],\\
\omega_R^{\rm Schw}&=\frac{\sqrt{\ell(\ell+1)}}{3\sqrt{3}M},\\
\omega_I^{\rm Schw}&=\frac{1}{3\sqrt{3}M}\sqrt{\left(n+\frac{1}{2}\right)}.
\end{align}
For the corrected Schwarzschild background considered here, we obtain
\begin{align}
\frac{\delta\omega_R}{\omega_R^{\rm Schw}}&=\frac{540}{19683}\frac{\gamma}{M^4}+\frac{832}{19683}\frac{\eta}{M^6},\\
\frac{\delta\omega_I}{\omega_I^{\rm Schw}}&=\frac{1404}{19683}\frac{\gamma}{M^4}+\frac{7232}{19683}\frac{\eta}{M^6},
\end{align}
where $\delta\omega_{R}$ ($\delta\omega_{I}$) is the EFT correction to the real (imaginary) part of the frequency, respectively.

In summary, we have computed the QNM frequencies in the eikonal limit and shown that they are determined entirely by the properties of the unstable null orbit at the photon sphere. Higher-curvature corrections modify these quantities through shifts in the photon sphere location, leading to corresponding corrections to both the real and imaginary parts of the QNM spectrum.

\section{Weak-Field Gravitational Lensing in Effective Field Theory}
\label{sec:weak lens}

To confront the theory with observations, it is essential to study gravitational lensing in both the weak- and strong-field regimes. These regimes probe different regions of the spacetime geometry and provide complementary information about the underlying gravitational theory.

In this section, we focus on the weak deflection limit, where light rays pass far from the compact object and the deflection angle can be treated perturbatively. In the strong deflection limit, light rays approach the photon sphere, where nonlinear effects become dominant, leading to large deflection angles and the formation of relativistic images. The analysis of the strong-field regime is deferred to the next section. In the following, we derive the corresponding observables in the weak-field regime and analyze how they are affected by higher-curvature corrections.

We consider a general static and spherically symmetric spacetime
\begin{align}
ds^2 = -A(r)\,dt^2 + B(r)\,dr^2 + C(r)\,d\Omega^2,
\end{align}
where $A(r)$, $B(r)$, and $C(r)$ are arbitrary functions of $r$.

For null geodesics ($ds^2=0$), restricting to the equatorial plane $\theta=\pi/2$,
we have
\begin{align}
0 = -A(r)\dot{t}^2 + B(r)\dot{r}^2 + C(r)\dot{\phi}^2.
\end{align}
The conserved energy $E$ and angular momentum $L$ are
\begin{align}
E = A(r)\dot{t}, \qquad L = C(r)\dot{\phi}.
\end{align}
Using these, the radial equation becomes
\begin{align}
\dot{r}^2 = \frac{E^2}{A(r)B(r)} \left[ 1 - \frac{b^2A(r)}{C(r)} \right],
\end{align}
where $r_0$ is the distance of closest approach, defined by $\dot{r}=0$. The condition $\dot{r}=0$ then leads to
\begin{align}
\frac{b^2 A(r_0)}{C(r_0)} = 1.
\end{align}

The deflection angle is then given by
\begin{align}
\label{deflection angle}
\alpha = 2 \int_{r_0}^{\infty} \sqrt{\frac{B(r)}{C(r)}}
\left[
\frac{C(r)}{C(r_0)} \frac{A(r_0)}{A(r)} - 1
\right]^{-1/2}dr - \pi.
\end{align}

We now assume the weak-field limit, in which the metric functions can be expanded as
\begin{align}
A(r) = 1 + \lambda\, a(r), \quad
B(r) = 1 + \lambda\, b(r), \quad
C(r) = r^2 \left[1 + \lambda\, c(r)\right],
\end{align}
with $\lambda \ll 1$.

We also assume that the closest approach distance satisfies $r_0 \simeq b$,
where $b$ is the impact parameter.

Expanding the integrand to first order in $\lambda$, we obtain
\begin{align}
\alpha&\sim2\int_{r_0}^{\infty}
\frac{r_0 \, dr}{r \sqrt{r^2 - r_0^2}}
\left[
1
+ \frac{\lambda}{2}\bigl(b(r) - c(r)\bigr)
- \frac{\lambda}{2}
\frac{r^2}{r^2 - r_0^2}
\bigl(c(r) - c(r_0) + a(r_0) - a(r)\bigr)
\right]-\pi\nonumber\\
&=2\lambda\int_{r_0}^{\infty}
\frac{r_0 \, dr}{r \sqrt{r^2 - r_0^2}}
\left[\frac{1}{2}\bigl(b(r) - c(r)\bigr)
- \frac{1}{2}
\frac{r^2}{r^2 - r_0^2}
\bigl(c(r) - c(r_0) + a(r_0) - a(r)\bigr)
\right]
\end{align}

Solving the equations of motion, we obtain the perturbation functions as
\begin{align}
a(r) &=-\frac{2M}{r}-\frac{1}{r^{10}}\Big(64\eta M^{3} (22M - 16 r)- 40\gamma M^{3}r^3\Bigr)\nonumber\\
&\sim-\frac{2M}{r}+\frac{8M^3}{r^{9}}\Big(128\eta+5\gamma r^2\Bigr),\\
b(r) &=\frac{2M}{r}+\frac{24M^{2}r}{(r-2M)^2}\left(\frac{1072 \eta M^{2}}{3 r^{9}}- \frac{192 \eta M}{r^{8}}+ \frac{49 \gamma M}{3 r^{6}}- \frac{9 \gamma}{r^{5}}\right)\nonumber\\
&\sim\frac{2M}{r}-\frac{24M^{2}}{r^9}\left(192 \eta M+9 \gamma r^3\right),\\
c(r) &=0,
\end{align}
which will be used in the following analysis. Substituting these expressions into the deflection angle, we obtain
\begin{align}
\alpha=2\int_{r_0}^{\infty}
\frac{r_0 \, dr}{r \sqrt{r^2 - r_0^2}}
\left[\frac{1}{2}b(r)
- \frac{1}{2}\frac{r^2}{r^2 - r_0^2}\bigl(a(r_0) - a(r)\bigr)\right].
\end{align}
We first extract the leading contribution in general relativity and then evaluate the leading corrections arising from higher-curvature terms by expanding the integrand accordingly.
\begin{align}
\alpha&\sim2\int_{r_0}^{\infty}
\frac{r_0 \, dr}{r \sqrt{r^2 - r_0^2}}
\Bigl[\frac{M}{r}-\frac{12M^{2}}{r^9}\left(192 \eta M+ 9 \gamma r^3\right)\nonumber\\
&\ \ \ \ \ \ \ \ \ \ +\frac{1}{2}\frac{r^2}{r^2 - r_0^2}\Bigl(\frac{2M(r-r_0)}{r_0r}-\frac{1024\eta M^3(r^9-r_0^9)}{r_0^{9}r^9}-\frac{40\gamma M^{3}(r^7-r_0^7)}{r_0^7r^7}\Bigr)\Bigr]\nonumber\\
&=\frac{4M}{r_0}-\frac{196608M^3\eta}{35r_0^9}-\frac{128M^3\gamma}{r_0^7}-\frac{135\pi M^2\gamma}{4r_0^6},
\end{align}
where the first term reproduces the standard general relativistic result, while the remaining terms represent corrections induced by higher-curvature effects.

The weak-lensing results provide a useful comparison with the standard general-relativistic expansion. In the weak-field regime, the deflection angle admits an expansion in powers of $M/r_0$, with the leading contribution given by the standard Einstein term and the next-to-leading contribution arising at $O(M^2/r_0^2)$. The higher-curvature EFT corrections obtained in this work are found to be comparable to or smaller than this second-order GR contribution for the parameter range considered. Therefore, the EFT corrections are not expected to dominate the weak-lensing signal and may be difficult to distinguish from higher-order general-relativistic effects in this regime. The main role of the weak-lensing analysis in the present work is consequently to characterize the behavior of the EFT corrections in the weak-field regime and to contrast it with the stronger modifications that arise in the vicinity of the photon sphere.

It is useful to compare the magnitude of the EFT correction with the standard higher-order terms in general relativity. For a Schwarzschild black hole, the weak-deflection angle is given by
\begin{align}
\alpha_{\rm GR}(r_0)
={}&
\frac{4M}{r_0}
+\left(\frac{15\pi}{4}-4\right)\frac{M^2}{r_0^2}
+\left(\frac{122}{3}-\frac{15\pi}{2}\right)\frac{M^3}{r_0^3}
\nonumber\\
&+\left(\frac{3465\pi}{64}-130\right)\frac{M^4}{r_0^4}
+\left(\frac{7783}{10}-\frac{3465\pi}{16}\right)\frac{M^5}{r_0^5}
\nonumber\\
&+\left(\frac{310695\pi}{256}-\frac{21397}{6}\right)\frac{M^6}{r_0^6}
+O\left(\frac{M^7}{r_0^7}\right),
\end{align}
in Refs.~\cite{Keeton:2005jd,Chiba:2017nml}. The EFT corrections obtained in the present analysis are found to be comparable to or smaller than the higher-order corrections in general relativity for the parameter range considered. This indicates that, in the weak-field regime, the effects of higher-curvature corrections are subleading relative to the leading GR contribution and can be comparable to the intrinsic higher-order corrections already present in general relativity. Thus, the weak-lensing results primarily serve to characterize the weak-field behavior of the EFT corrections and to provide a comparison with the strong-field regime, where the photon-sphere geometry plays a dominant role.

The leading contribution from higher-curvature corrections may enter at the same order as subleading terms in general relativity. As a result, these effects cannot be cleanly disentangled from higher-order general relativistic contributions. Therefore, the impact of higher-curvature terms should not be regarded as an independent leading effect, but rather can be assessed by identifying deviations from the predictions of general relativity.

\section{Strong-Field Gravitational Lensing in Effective Field Theory}
\label{sec:strong lens}

In the previous section, we analyzed gravitational lensing in the weak-field regime, where the deflection angle can be treated perturbatively. In this section, we turn to the strong-field regime, where light rays propagate in the vicinity of the photon sphere and nonlinear effects become significant, leading to large deflection angles and the formation of relativistic images.

\subsection{Divergent term of the deflection angle}

In this subsection, we review the strong deflection limit analysis of gravitational lensing developed by Valerio Bozza \cite{Bozza:2002zj}. We define new variable
\begin{align}
z = \frac{A(r) - A_0}{1 - A_0},
\label{z variable}
\end{align}
where $A_0 = A(r_0)$. The integral (\ref{deflection angle}) in the deflection angle becomes
\begin{align}
I(r_0) = \int_{0}^{1} R(z,r_0)\, f(z,r_0)\, dz,
\end{align}
with
\begin{align}
R(z,r_0) &= \frac{2 \sqrt{AB}}{C A'} (1 - A_0)\, \sqrt{C_0}, \\
f(z,r_0) &= \frac{1}{\sqrt{A_0 - \left[(1 - A_0) z + A_0\right] \frac{C_0}{C}}}.
\end{align}
Here, all functions with the subscript $0$ are evaluated at $r_0$ and ${}^\prime$ denotes differentiation with respect to $r$.

The function $R(z, r_0)$ is regular for all values of $z$ and $r_0$, whereas $f(z, r_0)$ diverges in the limit $z \to 0$. To determine the leading behavior of this divergence in the integrand, we expand the argument of the square root in $f(z, r_0)$ up to second order in $z$:
\begin{align}
f(z,r_0) \sim f_0(z,r_0) = \frac{1}{\sqrt{p(r_0) z + q(r_0) z^2}},
\end{align}
where
\begin{align}
p(r_0) &= \frac{1 - A_0}{C_0 A'_0} \left(C'_0 A_0 - C_0 A'_0\right), \\
q(r_0) &= \frac{(1 -A_0)^2}{2 C_0^2 A'_0{}^3}
\left[
2 C_0 C'_0 A'_0{}^2
+ \left(C_0 C''_0 - 2 C'_0{}^2\right) A_0 A'_0
- C_0 C'_0 A_0 A''_0
\right].
\end{align}

When $p(r_0)$ remains nonvanishing, the leading divergence of $f_0$ scales as $z^{-1/2}$, which is integrable and therefore leads to a finite contribution. In contrast, if $p(r_0)$ vanishes, the divergence instead scales as $z^{-1}$, causing the integral to diverge.

From the structure of $p(r_0)$, one finds that it becomes zero at $r_0 = r_{ph}$, where $r_{ph}$ is specified by the condition introduced in the previous section:
\begin{align}
\label{ph radius}
\frac{d}{dr}\left(\frac{A(r)}{C(r)}\right)\bigg|_{r=r_{ph}} = 0.
\end{align}
At this location, the impact parameter reaches its critical value $b_c$, and photons follow unstable circular orbits around the black hole. For trajectories with $r_0 < r_{ph}$, photons are captured by the central object and do not escape to infinity.

To evaluate the integral, we decompose it into two contributions:
\begin{align}
I(r_0) = I_D(r_0) + I_R(r_0),
\end{align}
where
\begin{align}
I_D(r_0) &= \int_{0}^{1} R(0,r_{ph})\, f_0(z,r_0)\, dz,
\end{align}
captures the divergent behavior, and
\begin{align}
I_R(r_0) &= \int_{0}^{1} g(z,r_0)\, dz,
\end{align}
with
\begin{align}
g(z,r_0) = R(z,r_0) f(z,r_0) - R(0,r_m) f_0(z,r_0),
\end{align}
represents the original integrand with the singular contribution removed.

We proceed by computing each part independently and then combining the results to reconstruct the deflection angle. In this subsection, we focus on $I_D$ and its divergence, while in the following subsection we now examine how the finite contribution is obtained.

The integral $I_D(r_0)$ admits an exact evaluation:
\begin{align}
I_D(r_0) = \frac{R(0,r_{ph})}{2\sqrt{q(r_0)}} 
\log \left(\frac{\sqrt{q(r_0)} + \sqrt{p(r_0) + q(r_0)}}{\sqrt{p(r_0)}}\right).
\end{align}

where
\begin{align}
q(r_{ph}) = q(r_0)\big|_{r_0 = r_{ph}}= \frac{C_{ph} (1 - A_{ph})^2 \left(C''_{ph} A_{ph} - C_{ph} A''(r_{ph})\right)}{2 A_{ph}^2 C'_{ph}{}^2}.
\end{align}
Here, all functions with the subscript $ph$ are evaluated at $r_{ph}$.

Substituting these expressions into $I_D(r_0)$ and reorganizing terms, we obtain
\begin{align}
I_D(r_0)= - a \log\left( \frac{r_0}{r_{ph}} - 1 \right)+ b_D+ \mathcal{O}((r_0 - r_{ph})\log(r_0-r_{ph})),
\end{align}
with
\begin{align}
a &= \frac{R(0,r_{ph})}{\sqrt{q(r_{ph})}}, \\
b_D &= \frac{R(0,r_{ph})}{\sqrt{q(r_{ph})}}
\log \left( \frac{2(1 - A_{ph})}{A'_{ph} r_{ph}} \right).
\end{align}

This result provides the leading divergent contribution to the deflection angle, which is logarithmic in nature, as expected.

To determine the full coefficient $b$ relevant for strong-field gravitational lensing, following the method developed by Bozza \cite{Bozza:2002zj}, we must supplement $b_D$ with the corresponding contribution from the regular part of the integral defined in Eq.~(\ref{deflection angle}).

We expand $I_R(r_0)$ in powers of $(r_0 - r_{ph})$:
\begin{align}
I_R(r_0)
=
\sum_{n=0}^{\infty}
\frac{1}{n!} (r_0 - r_{ph})^n
\int_{0}^{1}
\left.
\frac{\partial^n g}{\partial r_0^n}
\right|_{r_0 = r_{ph}}
dz.
\end{align}

Without subtracting the divergent part from $R(z,r_0)f(z,r_0)$, the coefficient corresponding to $n=0$ would diverge, whereas all higher-order terms would be finite. By construction, however, the function $g(z,r_0)$ is regular at $z=0$ and $r_0 = r_{ph}$, as can be verified through a series expansion, using $p(r_{ph}) = 0$.

Restricting ourselves to terms up to $\mathcal{O}(r_0 - r_{ph})$, we keep only the leading term:
\begin{align}
I_R(r_0)
=
\int_{0}^{1} g(z,r_{ph})\, dz
+ \mathcal{O}(r_0 - x_{ph}),
\end{align}
and define
\begin{align}
b_R = I_R(x_{ph}).
\end{align}

Including also the $-\pi$ contribution in the deflection angle, we arrive at
\begin{align}
b = -\pi + b_D + b_R.
\end{align}

The quantity $b_R$ can be evaluated numerically for general metrics, since the integrand is regular. In several cases, however, an analytic expression can be obtained. For instance, in the Schwarzschild spacetime, the integral can be computed exactly. More broadly, one may expand the integral around the Schwarzschild limit in terms of metric parameters, allowing each term in the expansion to be determined separately.

In the general formula for the deflection angle, the impact parameter $u$ is introduced. We begin by defining it as a function of the closest approach distance $r_0$:
\begin{align}
\label{impact para}
u = \sqrt{\frac{C_{0}}{A_{0}}}.
\end{align}

From Eq.~(\ref{ph radius}), the minimum impact parameter is given by
\begin{align}
u_{ph} = \sqrt{\frac{C_{ph}}{A_{ph}}}.
\end{align}

Expanding Eq.~(\ref{impact para}) yields
\begin{align}
u - u_{ph}= c (r_0 - r_{ph})^2+\mathcal{O}\left( (r_0 - r_{ph})^3\right),
\end{align}
where
\begin{align}
c
&=
\frac{C''_{ph} A_{ph} - C_{ph} A''_{ph}}{4 \sqrt{A_{ph}^3 C_{ph}}}
=
q(r_{ph}) \sqrt{\frac{A_{ph}}{C_{ph}^3}} \frac{C_{ph}'^2}{2(1 - A_{ph})^2}.
\end{align}

Using this relation, the deflection angle may be expressed as a function of the angular position $\theta$:
\begin{align}
\alpha(\theta)
=
- a \log\left( \frac{\theta D_{OL}}{u_{ph}} - 1 \right)
+ \bar{b},
\end{align}
with
\begin{align}
\bar{a} &= \frac{a}{2} = \frac{R(0,x_{ph})}{2\sqrt{q(r_{ph})}}, \\
\bar{b} &= b + \frac{a}{2} \log \left( \frac{c r_{ph}^2}{u_{ph}} \right)
= -\pi + b_R + \bar{a} \log \left( \frac{2q(r_{ph})}{A_{ph}} \right).
\end{align}

\subsection{Strong Deflection Analysis with $2M=1$}

In this subsection, we analyze gravitational lensing in the strong deflection regime, taking into account the contributions from higher-curvature corrections. To this end, we adopt a convenient normalization of the radial coordinate. Throughout this section, we set $2M = 1$, where $M$ is the mass parameter of the central object. In this system of units, the Schwarzschild radius is located at $r = 1$, which simplifies the analytical expressions without loss of generality. The EFT couplings $\gamma$ and $\eta$ are dimensionful parameters. To work with dimensionless expansion parameters, we normalize them by appropriate powers of the black-hole mass scale. In particular, before setting $2M=1$, we define
\begin{align}
\hat{\gamma} \equiv \frac{\gamma}{(2M)^4},
\qquad
\hat{\eta} \equiv \frac{\eta}{(2M)^6}.
\end{align}
In the numerical analysis, we set $2M=1$, so that the parameters denoted by $\gamma$ and $\eta$ below should be understood as these dimensionless quantities.

The perturbative expansion requires not only $\hat{\gamma},\hat{\eta}\ll1$, but also that the actual higher-curvature corrections remain small compared with the leading-order terms. Since the perturbative solutions contain numerical coefficients that can be relatively large, the effective expansion parameters are determined by combinations such as $C_\gamma\hat{\gamma}$ and $C_\eta\hat{\eta}$, where $C_\gamma$ and $C_\eta$ denote the corresponding numerical coefficients. Therefore, the practical regime of validity is more restrictive than the naive conditions $\hat{\gamma}\ll1$ and $\hat{\eta}\ll1$. Throughout this work, we restrict our analysis to parameter values for which the higher-curvature corrections remain perturbatively small.

We consider null geodesics in a static, spherically symmetric spacetime described by the metric
\begin{align}
A(r)&=\left(1-\frac{1}{r}\right)-\frac{1}{r^{10}}\Big(8\eta(11- 16 r)- 5\gamma r^3\Bigr),\\
B(r)&=\frac{1}{\left(1-\frac{1}{r}\right)}+\frac{6}{\left(1-\frac{1}{r}\right)^2}\left(\frac{268 \eta}{3 r^{10}}- \frac{96 \eta }{r^{9}}+ \frac{49 \gamma }{6 r^{7}}- \frac{9 \gamma}{r^{6}}\right),\\
C(r)&=r^2.
\end{align}
These metric functions incorporate the higher-curvature corrections up to the order considered in the effective field theory expansion.

In this background, we evaluate the deflection angle in the strong deflection limit following the standard formalism. The relevant functions $R(z,r_{ph})$ and $f(z,r_{\rm ph})$ appearing in the integral expression of the deflection angle are given by
\begin{align}
R(z,r_{ph})&=2-\frac{8192(1-z)^9\left(2716254\eta-\frac{4251528\eta}{1-z}+\frac{255879\eta}{(1-z)^9}-\frac{531441\gamma}{8(1-z)^3}+\frac{2657205\gamma}{64(1-z)^9}\right)}{387420489},\\
f(z,r_{ph})&=(z^2 - \tfrac{2}{3}z^3)^{-1/2}
+ \frac{64}{59049}\, z^2 (z^2 -\frac{2}{3}z^3)^{-3/2}
\Big(
11712 \eta + 1215 \gamma
- 7040 \eta z
- 46080 \eta z^2
- 5670 \gamma z^2\nonumber\\
&\ \ \ \ + 118656 \eta z^3
+ 11340 \gamma z^3
- 126336 \eta z^4
- 11340 \gamma z^4
+ 42624 \eta z^5
+ 6480 \gamma z^5
+ 48960 \eta z^6
- 2025 \gamma z^6
\nonumber\\
&\ \ \ \ - 70720 \eta z^7+ 270 \gamma z^7
+ 40512 \eta z^8
- 11712 \eta z^9
+ 1408 \eta z^{10}
\Big),
\end{align}
where $z$ is defined as Eq.~(\ref{z variable}).

Substituting these expressions into the deflection angle formula and performing the integration, we obtain the regular part of the deflection angle as
\begin{align}
b_R=&2\ln[6(2-\sqrt{3})]
+ \Bigg[
\frac{16384}{1674728055}\,\eta
\left(
-18678783 + 4129595 \sqrt{3}
+ 5360355 \log(12)
- 10720710 \log\!\left(1+\sqrt{3}\right)
\right)\nonumber\\
&+ \frac{256}{280665}\,\gamma
\left(
-9924 + 2747 \sqrt{3}
+ 3465 \log(12)
- 6930 \log\!\left(1+\sqrt{3}\right)
\right)
\Bigg]\,
\end{align}

We next determine the strong deflection limit coefficients and the photon sphere quantities. The corresponding expressions are given by
\begin{align}
q_{ph}&=1-\frac{128(3904\eta+405\gamma)}{19683},\\
a\ &=2+\frac{256(448\eta+27\gamma)}{2187},\\
b_D&=\log[4]+\frac{256(992\eta+135\gamma+1344\eta\log[2]+27\gamma\log[8])}{6561},\\
u_{ph}&=\frac{3\sqrt{3}}{2}-\frac{32(832\eta+135\gamma)}{2187\sqrt{3}}.
\end{align}

Combining the divergent and regular contributions, we obtain the coefficient $\bar{b}$ appearing in the deflection angle as
\begin{align}
\bar{b}=-\pi+b_R+\log[6]+\frac{128(-640\eta-45\gamma+1344\log[6]\eta+81\log[6]\gamma)}{6561},
\end{align}

Finally, the deflection angle in the strong deflection limit can be written in terms of the angular position $\theta$ as
\begin{align}
\alpha(\theta)=&-\left(1+\frac{128(448\eta+27\gamma)}{2187}\right)\log\left(\frac{\theta D_{OL}}{\frac{3\sqrt{3}}{2}-\frac{32(832\eta+135\gamma)}{2187\sqrt{3}}}-1\right)+\log\left[216(7-4\sqrt{3})\right]\nonumber\\
&+\frac{128}{1674728055}\Bigg[64 \eta \left(-39910116 + 8259190 \sqrt{3}+ 5360355 \log(6)+ 10720710 \log(12)
- 21441420 \log\!\left(1+\sqrt{3}\right)\right)\nonumber\\
&+ 5967 \gamma \left(-21773 + 5494 \sqrt{3}+ 3465 \log(6)+ 6930 \log(12)- 13860 \log\!\left(1+\sqrt{3}\right)\right)\Bigg]-\pi.
\end{align}

We next evaluate the directly observable quantities associated with the strong-field lensing regime. In addition to the photon-sphere radius and the strong-field coefficients $(\bar a,\bar b)$, we consider the asymptotic image position $\theta_\infty$, the angular separation $s$ between the first relativistic image and e flux ratio between the first relativistic image and the remaining relativistic images. $\theta_\infty$ represents the asymptotic angular position at which the relativistic images accumulate, $s$ denotes the angular separation between the first relativistic image and $\theta_\infty$, and $\mathcal R$ is the flux ratio between the first relativistic image and the remaining relativistic images. For a distant observer, the angular shadow radius is approximately equal to the asymptotic image position, $\theta_{\rm sh}\simeq\theta_\infty$.

These quantities are determined from the critical impact parameter and the strong-deflection coefficients as
\begin{align}
\theta_\infty=\frac{u_c}{D_{OL}},
\qquad
s=\theta_\infty
\exp\left(\frac{\bar b-2\pi}{\bar a}\right),
\qquad
\mathcal R=
\exp\left(\frac{2\pi}{\bar a}\right).
\end{align}

For the present model, the resulting observable quantities are
\begin{align}
\theta_\infty&=\frac{\frac{3\sqrt{3}M}{2}-\frac{32(832\eta+135\gamma)}{2187\sqrt{3}}}{D_{OL}},\\
s&=\frac{(7-4\sqrt{3})e^{-3\pi}}{D_{OL}}\Biggl[216 +\frac{1}{620269651024}\Bigl(
-1074216839424\eta
+528588160\sqrt{3}\,\eta
-75481195491\gamma
+32782698\sqrt{3}\,\gamma
\nonumber\\
&\quad
+1029188160\pi\eta
+62026965\pi\gamma
+1372250880\eta\operatorname{arctanh}(1/\sqrt{3})
+82702620\gamma\operatorname{arctanh}(1/\sqrt{3})
\nonumber\\
&\quad
+2250834505920\eta\log 6
+135652972455\gamma\log 6
+686125440\eta\log 12
+41351310\gamma\log 12
\nonumber\\
&\quad
-343062720\eta\log 216
-20675655\gamma\log 216
-1372250880\eta\log(1+\sqrt{3})
-82702620\gamma\log(1+\sqrt{3})
\Bigr)\Biggr],\\
\mathcal R&=e^{2\pi}\left(1-\frac{256\pi(448\eta+27\gamma)}{2187}\right).
\end{align}

The relative fractional deviation from the Schwarzschild value are therefore
\begin{align}
\delta\theta_\infty
&=\frac{\theta_\infty-\theta_\infty^{\rm Sch}}{\theta_\infty^{\rm Sch}}\\
&=-\frac{64(832\eta+135\gamma)}{19683},\notag\\
\delta s&=\frac{s-s^{\rm Sch}}{s^{\rm Sch}}\notag\\
&=\frac{1}{133978244621184}\Bigl(
-1074216839424\eta
+528588160\sqrt{3}\,\eta
-75481195491\gamma
+32782698\sqrt{3}\,\gamma
\nonumber\\
&\quad
+1029188160\pi\eta
+62026965\pi\gamma
+1372250880\eta\operatorname{arctanh}(1/\sqrt{3})
+82702620\gamma\operatorname{arctanh}(1/\sqrt{3})
\nonumber\\
&\quad
+2250834505920\eta\log 6
+135652972455\gamma\log 6
+686125440\eta\log 12
+41351310\gamma\log 12
\nonumber\\
&\quad
-343062720\eta\log 216
-20675655\gamma\log 216
-1372250880\eta\log(1+\sqrt{3})
-82702620\gamma\log(1+\sqrt{3})
\Bigr),\\
\delta\mathcal R&=\frac{\mathcal R-\mathcal R^{\rm Sch}}{\mathcal R^{\rm Sch}}\notag\\
&=-\frac{256\pi(448\eta+27\gamma)}{2187}.
\end{align}

In this section, we apply the strong deflection expansion to the present setup and compute the corresponding coefficients characterizing the logarithmic behavior of the deflection angle. We then use these coefficients to evaluate the associated observable quantities, including the asymptotic image position and the image separation.

\section{Conclusion and Discussion}
\label{sec:conclusion}

In this paper, we have analyzed quasinormal modes (QNMs), weak gravitational lensing, and strong gravitational lensing in static, spherically symmetric spacetimes with higher-curvature corrections. By focusing on the role of the photon sphere, we have examined the geometric properties of unstable null geodesics and their relation to strong-field gravitational lensing and test-field QNMs, while weak gravitational lensing provides complementary information in the weak-field regime. The QNM analysis is restricted to the test-field limit, and gravitational perturbations of the metric are not considered.

We have shown that higher-curvature corrections within an effective field theory (EFT) framework modify key quantities associated with the photon sphere, including the photon sphere radius, the critical impact parameter, and the coefficients governing the strong deflection expansion. These quantities exhibit characteristic dependences on the EFT couplings and therefore provide a systematic characterization of how higher-curvature corrections modify the surrounding spacetime geometry.

Our analysis further demonstrates that the strong-deflection behavior of null geodesics is directly affected by the geometry near the photon sphere. In particular, the logarithmic divergence of the deflection angle and the associated strong-deflection coefficients receive corrections determined by the higher-curvature terms in the EFT. We have also examined the corresponding modifications of the eikonal QNMs of a test scalar field, allowing the effects of the EFT corrections on null-geodesic dynamics and test-field perturbations to be compared within the same background geometry. These results illustrate how different physical quantities respond to the same higher-curvature modifications of the spacetime.

An important outcome of this work is therefore a systematic description of the relation between higher-curvature corrections, photon-sphere geometry, strong gravitational lensing, and test-field QNMs. The photon sphere radius and critical impact parameter characterize the null-geodesic structure of the corrected spacetime, while the strong-deflection coefficients describe how this structure affects the propagation of light near the photon sphere. The corresponding test-field QNM corrections provide a complementary characterization of the same background geometry. We emphasize that this common geometric interpretation applies to the test-field analysis considered here and does not imply a direct correspondence between photon-sphere properties and gravitational QNMs in general higher-curvature theories.

Several directions for future work remain. First, it would be useful to apply the present framework to specific modified gravity models and to investigate how the EFT description is realized in concrete theories. In addition, while this work has focused on cubic and quartic Riemann tensor terms, which provide the relevant leading corrections in the vacuum background considered here, the analysis could be extended to more general higher-curvature theories. Related studies of gravitational lensing in modified theories of gravity have been presented in Refs. \cite{Badia:2017art,Jin:2020emq}. A detailed comparison with observational quantities, particularly those relevant to black-hole imaging, would also be necessary to determine the phenomenological significance of the corrections obtained here. Moreover, extending the analysis to rotating spacetimes would make the framework more applicable to realistic astrophysical systems. Finally, investigating higher-order terms in the strong-deflection expansion may further improve the theoretical description of gravitational lensing near the photon sphere.

Overall, our results provide a systematic characterization of how higher-curvature EFT corrections modify the photon-sphere geometry, strong gravitational lensing, and test-field QNMs in static, spherically symmetric spacetimes. The analysis establishes a common framework for comparing these effects at the level of the corrected background geometry and provides a basis for future studies connecting higher-curvature gravity with specific observational quantities.

\bibliography{references}

\end{document}